# INTERFERENCE-AWARE CHANNEL ASSIGNMENT FOR MAXIMIZING THROUGHPUT IN WMN

Vinay Kapse and Ms. U.N. Shrawankar

Department of Computer Science and Engineering, G.H. Raisoni College of Engineering, Nagpur, India
`vinaykapse82@gmail.com`
`urmilas@rediffmail.com`

## ABSTRACT

*Wireless Mesh network (WMN) is dynamically self-organizing and self-configured, with the nodes in the network automatically establishing an ad-hoc network and maintaining the mesh connectivity. The ability to use multiple-radios and multiple channels can be cashed to increase aggregate throughput of wireless mesh network. Thus the efficient use of available interfaces and channels without interference becomes the key factor. In this paper we propose interference aware clustered based channel assignment schemes which minimizes the interference and increases throughput. In our proposed scheme we have given priority to minimize interference from nearby mesh nodes in interference range than maximizing channel diversity. We simulated our proposed work using NS-3 and results show that our scheme improves network performance than BFSCA and Distributed Greedy CA.*

## 1. INTRODUCTION

Wireless Mesh Network (IEEE 802.11s) are dynamically self-organizing and self-configured, with the nodes in the network automatically establishing an ad-hoc network and maintaining the mesh connectivity. The wireless mesh network consists of mesh clients, mesh routers also called as mesh nodes and mesh gateways. Wireless mesh network has mesh clients such as laptops, smart phone and other wireless devices. The mesh routers are wireless routers that provide a routing functionality to and from wireless clients to internet gateways. Mesh routers also provide reliable, redundant and strong network backbone for providing internet services to mesh clients. Mesh gateways provide internet and high speed broadband connectivity to mesh clients through routers. Wireless mesh network have been subject of interest for research communities and wireless industries due to infrastructure less easy deployment, self organizing and self-configuring feature applicable in metropolitan areas. Through multi-hop communication, a large coverage area can be benefited by mesh routers with lower transmission power. Most of the routers have minimal mobility.

 Multi-Channel Multi-Radio communication, fixed mobility model in mesh network diversify the capabilities of ad-hoc networks. These features bring many advantages to WMNs, such as low up-front cost, easy network maintenance, robustness, reliable service coverage, etc. Therefore wireless mesh network is widely accepted in the traditional application scenarios such as broadband home networking, community networking, building automation, high speed metropolitan area networks, and enterprise networking.

Wireless mesh network are similar to in concept with mobile ad hoc network with some important differences and with less constraints as compared to ad hoc network which can be exploited. The main difference is that the nodes of WMN are not mobile or with negligible mobility. These avoid frequent topology changes and link failures. Topology changes are caused only if new nodes are added or due to node failures or power offing mesh routers for maintenance purposes. The traffic is always concentrated on the links originating and terminating to mesh gateways. More over flow characteristics do not change frequently. This characteristic can be used to optimize network traffic based on previous traffic statistics. All Mesh clients in wireless mesh network try to gain access to internet. Thus most of the traffic is directed to and from mesh gateways saturating channels at gateways. So allocating the non interfering channels at links near gateways can considerably increase network throughput.

The IEEE 802.11s standard is designed to work in compatibility with IEEE 802.11 a/b/g physical standards. The IEEE 802.11b has 12 overlapping channels (Channel 0..11) and 3 non-overlapping channels (CH-1, CH-6, CH-11).Where as IEEE 802.11a/g has 12 non-overlapping channels. These non-overlapping channels can be operated in the 2.4 GHz band in neighbourhood of each other without causing interference. Interference between the neighbourhood channels results in increase in end-to-end delay, increase in retransmission and hence decrease in overall throughput. The use of multiple channels proves to be good effort to decrease this interference. But the use of multiple channels in a single radio environment can lead to the considerably large channel switching time among channels and hence a delay in the transmission. Switching an interface from one channel to another incurs delay. For example, wireless NICs are currently available that support both IEEE 802.11a and IEEE 802.11b and can switch between the two bands, However with the currently available hardware, switching across bands incurs a large delay, but the switching delay is expected to reduce in the future. To overcome this use of multi-radio NICs are recommended to be used at mesh nodes of mesh network.

The rest of the paper is structured as follows. In Section II related work is discussed. In Section III we describe different network model for channel assignment. In Section IV we will show simulation results and finally we will conclude paper.

## 2. RELATED WORK

Many approaches were proposed in the past to increase capacity by reducing interference on wireless links. In one type of these approach, focused on the use of multiple non-overlapping channels over a single wireless network interface card [4]-[6]. This type of approach requires a fast and efficient algorithm to switch in between the channel. This approach fails to an inefficient because of the significant delay generated in switching the channels with the use of commodity hardware NICs. The delay generated can be of the order of milliseconds. Sometimes these are higher than the normal packet transmission time. Moreover the use of channel switching requires changing in MAC layer and hardware.

Subramanian et al [9] designed a centralized channel assignment algorithm in which nodes listens all available channels on its neighbourhood nodes for which listening node is in interference range of other nodes and assigns channels which minimizes the interference from the set of nodes within their interference range. The above approach considers the multi radios, which does not work when number of interfaces is limited.

Ramachandaran introduced multi-radio conflict graph model and a centralized CA algorithm [6]. They extended graph colouring problem to represent channel assignment problem as colouring the nodes in multi-radio conflict graph. This CA algorithm traverses multi-radio graph in bread first order and assigns channels in greedy manner and recommends utilization of a dedicated radio assigned to common channel in order to ensure network connectivity.

Ko et al.[10] propose a distributed channel assignment algorithm where each node can choose greedily a channel that minimizes its local objective function depending only on local information. Every node selects a channel that minimizes the sum of interference cost within its interference range. The advantage of this approach is that channel assignment can be achieved based on local information among nodes. However they don't consider number of interface cards per node.

Naveed [3] proposed cluster based interference aware CA that exploits multiple paths between mesh router and gateway. This method recommends the use of localized default channel in a cluster to broadcasting with minimum overhead. Dedicating an interface on each mesh node in the cluster poses heavy overhead.

Shin et al.[13] showed that finding a channel assignment for optimal performance is NP-hard. They presented the channel assignment scheme, which uses randomized channel assignment in a distributed manner while maintaining network connectivity. Channel assignment at NIC is done randomly.

One solution proposed by [11, 6, 12] reserves one channel as default channel on default NIC and other channels operating on non-default NIC for mesh connectivity within network. This ensures network connectivity, at same time increases overhead and delay.

One approach proposed by hyacinth is to assign channels through routing protocol. The protocol allocates channels in order to maximize channel diversity within flow. But it does not consider interflow interference during channel assignment.

Other approach is to use multiple radio and multiple channels without the requirement of channel switching [6-10]. The multiple WNICs allow simultaneous transmission and reception on different channels. Previous literature shows the use of maximum number of NIC per mesh router is limited to 3 as installing more than 3 NIC on commodity devices increases the collision. Therefore in our analysis we have limited the maximum number of NIC to be used per node to 2.But this approach needs the proper utilization of WNICs and channel assignment scheme in such a way to reduce interference among neighbouring nodes and maximize the throughput. This paper does analysis of algorithms and schemes used for channel assignment.

## 3. PROBLEM FORMULATION

### 3.1. Wireless mesh network Architecture

Wireless mesh network consist of fixed routers that provide a strong backbone to network to aggregate traffic and retransmit traffic to mesh gateways which in turn provides access to internet over a large coverage area. In, turn wireless mesh routing plays a role relaying nodes to and fro from mesh gateways forming multi-hop wireless mesh network. The gateways are interface to wired internetworking which contains infrastructure resources such as file servers and application servers. The link between gateway and the wired network is point-to-point IEEE 802.11 standard or IEEE 802.16.

Each wireless mesh router consists of multiple radios which can be tuned to any of 3 IEEE 802.11b non-overlapping channels or 12 IEEE 802.11a/g non-overlapping channels. For two nodes to have successfully communication, the two nodes should be in direct communication range of each other. Moreover the NICs of two mesh points should be tuned to same frequency.

The two nodes in the interfering range of each other can interfere with each other if they are tuned to same channel.

### 3.2. Transmission and Interference Model

Transmission and interference from nearby wireless mesh nodes can be described using two models. These are protocol model and physical model.

#### 3.2.1. Protocol Model

Let $Rt$ and $Ri$ denote the fixed transmission range and interference of all wireless interfaces respectively where $Ri > Rt$ (approximately $Ri = 2Rt$ $Ri = 2Rt$). Let distance $(u,v)$ represent the Euclidean distance between two nodes $u,v \in V$. For two nodes $u,v \in V$ direct communication is only possible if the distance $d(u,v) < Rt$ and at least one of the interfaces of the nodes operate in same channel. We assume that wireless links are symmetric that is if $u$ can transmit to $v$ than $v$ can also receive successful transmission from $u$. Two links $e1(u1,v1)$ and $e2(u2,v2)$ interfere with each other if both edges operate on a common channel and any of the distances $d(u1,u2), d(u1,v1), d(v1,u2), d(v1,v2) \leq Ri$.

#### 3.2.2. Physical Model

The transmission is successful if $SNRij$ (Signal to noise ratio) is greater than $SNRthres$ (threshold) where $SNRij$ denotes the signal-to-noise ratio at node $nj$ for transmission received from node $ni$.

#### 3.2.3. Channel Assignment problem formulation

The channel assignment problem is divided into two sub-problems. One is assigning interfaces to the virtual link between communicating nodes known as neighbour-to-interface binding. Second is assigning channels to interfaces known as interface-to-channel binding. The channel should be assigned to virtual link so that its available bandwidth should be proportional to the load it carries.

The goal of channel assignment is assigning channels to each node from set of non-overlapping channels such that the sum of loads on interfering link is minimized. The objective is to assign available interfaces on the nodes with the goal of minimizing the overall network interference i.e. minimizing interfering links.

CA problem is NP-Hard even with the knowledge of network topology and network traffic. NP-Hardness was proved by reducing multiple subset problems to CA problem [11]. It is also shown that minimum edge colouring is subset problem of CA problem [7]. Solution to the CA problem should address two important issues of wireless mesh networks: Connectivity and Interference. Connectivity changes in the network topology can cause problems by affecting network partitions, and affecting paths used by existing flows.

## 4. PROPOSED CA

We did some simulations on topology design for channel assignment to minimize interference and maximize throughput by giving priority to reduce interference due to hidden terminal problem rather than maximizing channel diversity. The simulations were performed for below 3 Topologies.

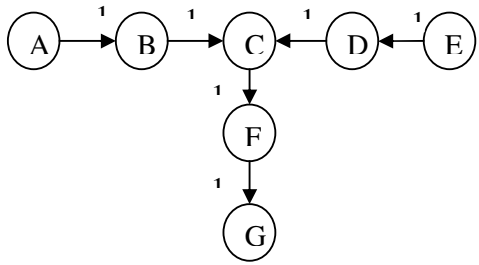
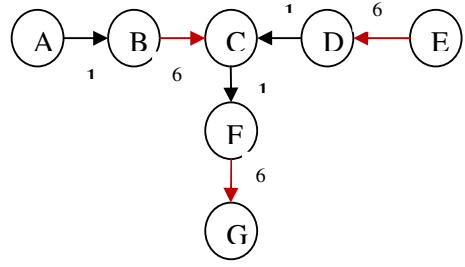

Figure 1. Single-Channel Single-Radio Topology

Figure 2. Multi-Radio Multi-Channel Topology

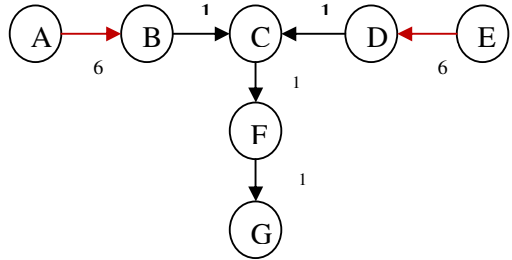

Figure 3. Interference-Aware multi-channel multi-radio topology design

We found from simulation results that increasing channel diversity does not affect much to aggregate throughput of network but minimizing interference considerably increases aggregate throughput of network.

### 4.1. Proposed Channel Assignment Scheme

In this section we explain in details of our proposed algorithm and the pseudo code. We propose a Distributed Channel Assignment scheme. Given a WMN topology graph proposed algorithm works in three phases. The first phase involves two parts. In the first part calculation of Euclidean distance between two wirelesses mesh points is calculated. In second part, formation of clusters takes place. In second phase interface allocation and channel assignment takes place in greedy fashion. In third phase Channel reassignment takes place considering interference from nearby nodes operating on same channel.

We say that mesh routers are placed in 3-dimensional space with x, y, z coordinates. The distance between two nodes will be calculated by Euclidean distance. If the distance $Dij$ between two mesh nodes is less than the transmission range $Xij$ than two nodes can receive and transmit data successfully without error and will result in formation of wireless link.

$$Dij = \sqrt{(x1-x)2 + (y1-y)2 + (z1-z)2}$$

Clustering method used in proposed CA is based on the technique proposed by Gonzalez [16]. The algorithm makes uniform r-hop clusters where r is the maximum hop distance from cluster head. In our implementation we selected r=2 because it represents interference domain size.

Initially clusters are formed by selecting mesh gateways as cluster heads. Each mesh node selects one of the cluster heads with minimum hop distance as its cluster head.

In the second phase interfaces are allocated to links. The binding is necessary to facilitate communication over a link by assigning the same channel to both interfaces assigned to link. The number of NICS required is equal to number of incident links on the mesh nodes. This is required to preserve connectivity within the network.

The channels are allocated to logical links in an greedy fashion, each type of channel is equally distributed all over the network. Distinct channels are assigned first to links incident on gateways. This makes sure that the load is balanced on the different links operating on different channels. The channels are assigned in way to provide maximum channel diversity.

In the third phase we find out the wireless links that operate in interference range of each other and operate on same channel. Our algorithm reassigns the channels on links in a way to minimize interference.

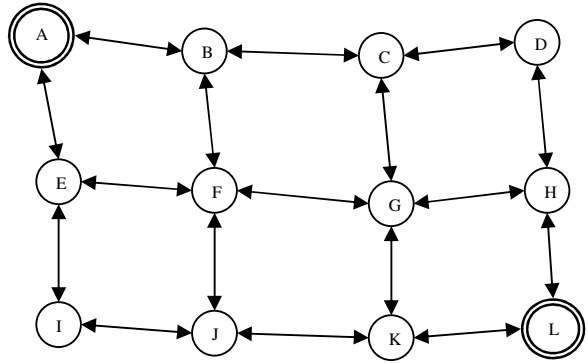

Figure 4. Mesh Grid Network and cluster creation

### 4.1. Pseudo Code

In this subsection we explain proposed algorithm pseudo code. First, we explain the notations used in our pseudo codes. Let $G<V, E>$ be the set of N nodes and $E \subseteq VxV$ be the set of $L$ links. Let $Vg \subseteq V$ is the set of gateway nodes. $M(u)$ represent the number of radio interfaces available on mesh router $u \in V$. Let $K$ be the set of orthogonal channels available in the network. $X$ represents the set of Clusters in the network.

The abstract pseudo code is as follows.

**Procedure** *ConstructLinks*(G)
  **begin**
1.    **for all** $vi \in V$ **do**
2.       for all $vj \in V$
3.         $Dij = \sqrt{(xi-xj)2 + (yi-yj)2 + (zi-zj)2}$
4.         if($Dij < Rt$)
5.           $E \leftarrow E \cup Eij$
6.         End if

7.   End for
8. End for
9. End *ConstructLinks()*

In *AllocateInterface*() procedure we calculate number of incident links on the mesh nodes. The mesh nodes are assigned number of NICs equal to number of incident links.

Procedure *AllocateInterface(G,E)*
 **begin**
1. **for all** $v_i \in V$ **do**
2.  **count = 0**
3.  **for all** $e_i \in E$ **do**
4.   **if**($\exists v_i \in V(e_i)$)
5.    count++
6.   **End if**
7.  **End for**
8.  $v_i = allocateNICs(count)$
9. **End for**
10. **End** *AllocateInterface()*

In ConstructCluster() one cluster is created per gateway node and gateway node acts as cluster headline. Mesh nodes joins appropriate cluster head depending on the minimum hop distance from the cluster head.
Procedure *ConstructCluster(G,Vg,HopCount,C)*
 **Begin**
1. **for all** $v \in Vg$ **do**
2.  $createCluster(Xv)$
3.  $X \leftarrow X \cup Xv$
4.  $CHID(v) \leftarrow Vg$
5. End for
6. for all $v \in \{V - Vg\}$ do
7.  $g \leftarrow selectMinHopGateway(HopCount)$
8.  $addClusterMember(Xg, v)$
9.  $CHID(v) \leftarrow g$
10. End for
11. for all $x \in X$ do
12.  if $(ClusterDist(x) > r)$
13.   $v \leftarrow selectMaxHopNode(x)$
14.   $i \leftarrow CHID(v)$
15.   $removeClusterMember(Xi, v)$
16.   $createCluster(X, Xv)$
17.   $X \leftarrow X \cup Xv$
18.   $CHID(v) \leftarrow v$
19.   for all $u \in \{V - Vg\}$ do
20.    $w \leftarrow CHID(u)$

21.             if $HopDist(u,w) \rangle HopDist(u,v)$
22.                 $deleteClusterMember(Xv,u)$
23.                 $addClusterMember(Xv,u)$
24.                 $CHID(u) \leftarrow v$
25.             end if
26.         end for
27.     end if
28.   end for
29. End *ConstructCluster()*

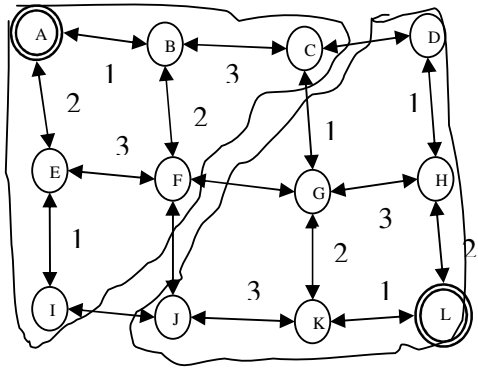
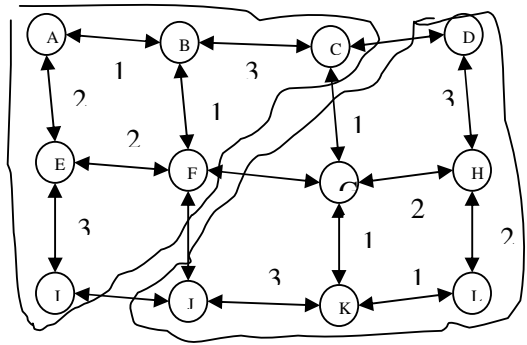

Fig. 5. Channel assignment within cluster

Fig 6. Interference-Aware Channel Re-assignment within cluster

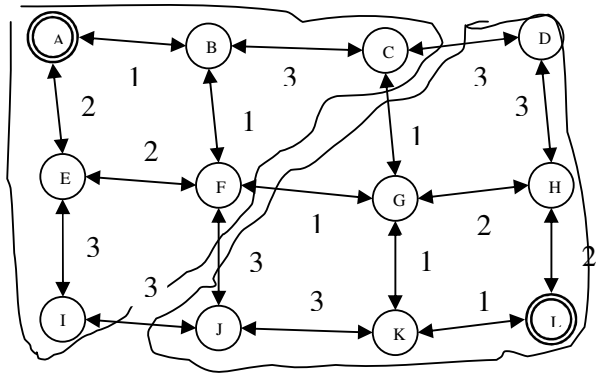

Fig. 7 Channel Assignment at border nodes

In second phase of channel assignment links at 3 hop distances is observed. If the links within 3 hop distance interfere with each other than channels are swapped with non-interfering channels. Thus in second phase we try to eliminate interfering links.

**Procedure** *ChannelReAssignment(G,X,E)*
   **Begin**

1. $Erassign = E$
2. $Etemp1 = \phi$
3. $Etemp2 = \phi$
4. **for all** $e \in Ereassign$ **do**
5.     **if** $(\exists vi \in V(ei))$
6.         $Etemp1 = Etemp1 \cup ei$
7.     End if
8.     for all $e1 \in Etemp1$
9.         if $(\exists v1i \in V(e1i))$
10.         $Etemp2 = Etemp2 \cup e1i$
11.         End if
12.     End for
13.     If $(Etemp2 \neq \phi)$
14.       $ci = getChannel(ei)$
15.       $c1i = getChannel(e1i)$
16.       $c2i = getChannel(e2i)$
17.       if $(ci = c2i)$
18.         $c2i = selectChannel(e2i)$
19.       End if
20.     End if
21.     $Erassign = Erassign - ei - e2i$
22.   End for
23. End Begin

## 4. SIMULATION RESULTS

Our proposed work was analyzed using NS-3 Simulator. Mesh network coverage on area 600m * 600m was established using fixed distribution of mesh router. Each mesh router are equipped first with four WNIC. The performance metrics will be obtained by averaging the results from thirty simulation runs for every experiment.

The network model was constructed with a propagation loss model of 50 db for the direct link between two nodes. The link is symmetric in nature. The distance between to communicating nodes is set 200m abroad. The propagation loss model of 200 db was assigned as a default loss model which implicitly means that there is no link between the two nodes or the two nodes are outside the reception range of each other. The channels used were non-overlapping channels (CH-34, CH-38 to CH-42) of IEEE 802.11a standard. The mobility of the mesh nodes is set to constant position mobility model. An UDPSocket were opened at the transmit end that generated packets of size 200 bytes at a data rate of 10kbps.This ensured the Constant Bit Rate (CBR) streams saturating the channels. The standard protocol stack containing HWMP for routing was used for all scenarios. The IEEE 802.11s link layer peer management link protocol was used. The positions of the mesh nodes are fixed and nodes are having zero mobility. The delay experienced by the packets to traverse from the transmitter to reception is kept constant. The maximum queue length at each interface is set to 255 packets. Total number of packets in queue will be used as a metric to estimate interference.

Simulation was run for 100 seconds. The .pcap files and trace files and flow monitor statistics was studied for analysis. The use of flow monitor was to collect flow statistics of every flow. Aggregate throughput of network was calculated as the summation of throughput for individual flows. The packet loss is calculated as the number of packets lost after time period of 10sec or dropped by receiving node due to interference, ttl timeout or invalid checksum.

TABLE I. INPUT PARAMETERS FOR SIMULATION

| Parameters | Values |
|---|---|
| Simulation Time | 100 Seconds |
| Simulation Area | 600m * 600m |
| Propogation Model | Two-ray Ground Reflection |
| Transmission range | 250 meter, 50db loss model |
| Traffic Type | CBR (UDP) |
| Packet Size | 128,256,1024,2048 bytes |
| Data Rates | 10 kbps |
| Number of nodes | 12 |
| Number of radios | 4 |
| Number of connections | 17 |
| Link layer max queue length | 255 |

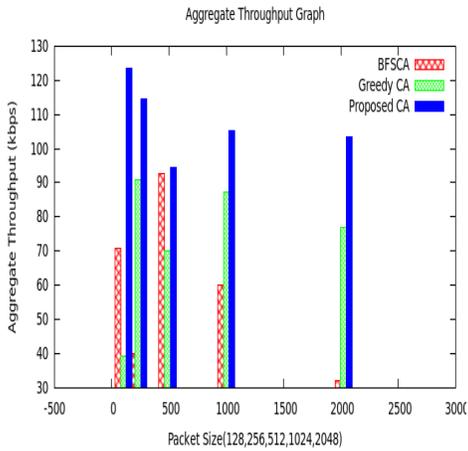

Figure 8. Aggregate Throughput Vs Packet Size (128,256,1024,2048)

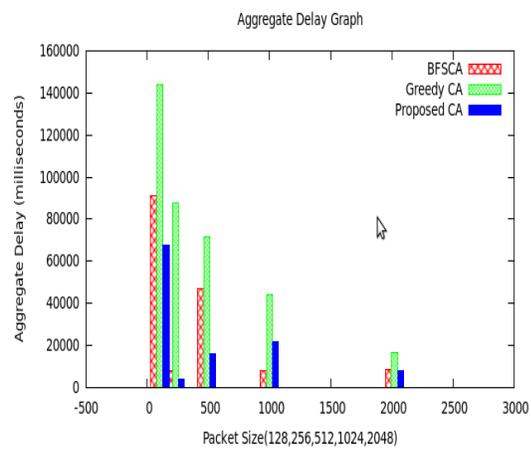

Figure 9. Aggregate Delay Vs Packet Size

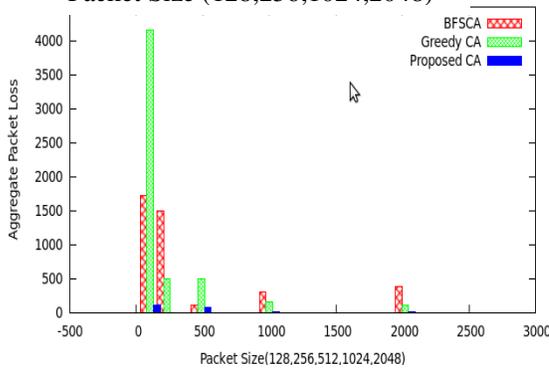

Figure 9. Aggregate Packet loss Vs Packet size

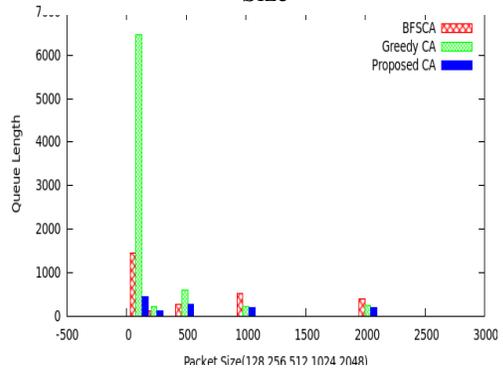

Figure 10. Aggregate Queue length Vs packet size

">International Journal on AdHoc Networking Systems (IJANS), Vol. 1, No. 1, June 2011

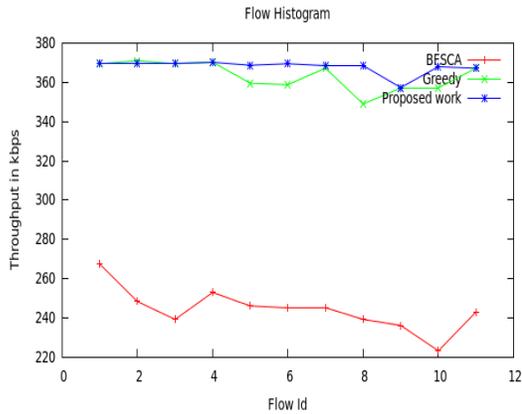
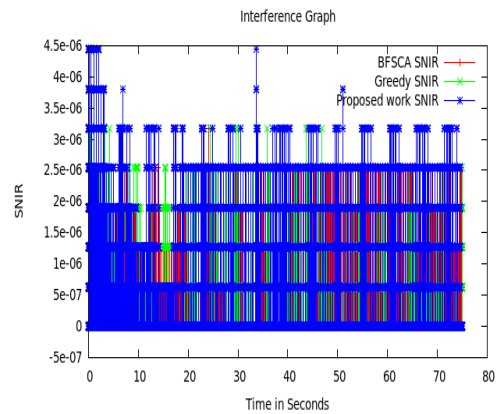

Figure 11. Flow Id Vs Throughput    Figure 12. SNIR vs time

From simulation results our proposed channel assignment scheme shows increase in throughput as compared to BFSCA. Initially less packet size results in smaller contention time. As the packet size increases from 128 bytes to 1024 bytes packets needs large transmission time which increases the contention time for interference between the flows that makes the property of collision between simultaneous transmissions becomes high due to many flows use the same channel to communicate with each other. Results show that our proposed heuristic performs better than BFSCA and Distributed Greedy Channel Assignment Schemes.

## 5. CONCLUSION

The Simulation were studied with respect to Aggregate throughput, Aggregate Delay Aggregate packet loss and Aggregate Queue length experienced by the mesh network when BFSCA, Greedy Channel Assignment and Our Proposed CA heuristic was operated. The BFSCA reserves one channel and one radio as an default channel as common to preserve connectivity. Though this BFSCA does preserve connectivity but one channel and one radio is always reserved which does not efficiently utilizes available resources. The Greedy channel assignment scheme selects the channels in greedy way. The less interfering channels are selected for channel assignment for wireless links. Our proposed channel assignment scheme allocates the channels that do not interfere with other channels in two hop distances. In doing so some of the nodes may be operating on one channel and one radio and other nodes may be inactive. This may lead to inefficient use of available resources in terms of channels and radio at nodes. This inefficient use is resources in our proposed work is acceptable since use of all channels and radios may increase interference and decrease throughput.

From the simulated results SNIR of our proposed work is high as compared to BFSCA and Greedy CA scheme. The average queue length was calculated at each interfaces of each node. From queue length results it is seen that the total interference experienced by each flow through our proposed channel assignment scheme is less than BFSCA and Greedy CA scheme. Aggregate throughput, Packet loss, Aggregate Delay experienced by simulating our proposed channel assignment scheme is better than BFSCA and Greedy CA.